\documentclass[doublecol]{epl2} 

\usepackage{amsmath}
\usepackage{amssymb,amsfonts,latexsym}
\usepackage{bm}
\usepackage[mathcal]{euscript}
\usepackage{graphicx}
\usepackage{epsfig}
\usepackage{color}
\usepackage{pifont}

\definecolor{blue}{rgb}{0,0,1}
\definecolor{bleuf}{rgb}{0,0,0.9}
\definecolor{rougef}{rgb}{0.9,0,0}
\definecolor{green}{rgb}{0,0.5,0}
\definecolor{red}{rgb}{1,0,0}
\definecolor{pink}{rgb}{0.9,0.3,0.7}
\definecolor{azur}{rgb}{0,0.5,0.5}
\definecolor{orange}{rgb}{1,0.5,0.2}
\definecolor{brown}{rgb}{0.5,0,0}

\newcommand{\be}{\begin{equation}}
\newcommand{\ee}{\end{equation}}
\newcommand{\ben}{\begin{equation*}}
\newcommand{\een}{\end{equation*}}
\newcommand{\ba}{\begin{eqnarray}}
\newcommand{\ea}{\end{eqnarray}}

\newcommand{\leg}[1]{\textbf{#1}}

\newcommand{\mytwofigures}[3][]{%
  \hbox to\hsize{%
    \vbox{%
       \hbox{\includegraphics*[#1]{#2}}%
    }%
    \hfill
    \vbox{%
       \hbox{\includegraphics*[#1]{#3}}%
}
    \hfill
  }%
}

\graphicspath{{figures/}}

\title{Dynamics of the Contacts reveals Widom Lines for Jamming}
\author{C. Coulais\inst{1} \and R. P. Behringer\inst{2} \and O. Dauchot\inst{3}}
\institute{                    
  \inst{1} SPHYNX/SPEC, CEA-Saclay, URA 2464 CNRS, 91 191 Gif-sur-Yvette, France\\
  \inst{2} Department of Physics and Center for Nonlinear and Complex Systems, Duke University, Durham, North Carolina 27708-0305, USA\\
  \inst{3} EC2M, ESPCI-ParisTech, UMR Gulliver 7083 CNRS, 75005 Paris, France
}

\abstract{
We experimentally study the vicinity of the Jamming transition by
investigating the statics and the dynamics of the contact network of
a horizontally shaken bi-disperse packing of photo-elastic discs.
Compressing the packing very slowly, while maintaining a mechanical
excitation, yields a granular glass, namely a frozen structure of
vibrating grains.  In this glass phase, we observe a remarkable
dynamics of the \emph{contact network}, which exhibits strong
dynamical heterogeneities. Such heterogeneities are maximum at a
packing fraction $\phi^*$, \emph{distinct} and smaller than the
structural packing fraction $\phi^{\dagger}$, which is indicated by an abrupt
variation of the average number of contact per particle.  We
demonstrate that the two cross-overs, one for the maximum dynamical
heterogeneity, and the other for static jamming, converge at point J
in the zero mechanical excitation limit, a behavior reminiscent of the
Widom lines in the supercritical phase of a second order critical
point. Our findings are discussed in the light of recent numerical and
theoretical studies of thermal soft spheres.}


\begin{document}

\maketitle
\vspace{-2mm} At large packing fraction, disordered packings of
particles with repulsive contact interactions jam into a rigid
state. For frictionless and a-thermal particles, the jamming
transition coincides with the onset of isostaticity and a number of
geometrical and mechanical quantities exhibit clear scaling laws with
the distance to jamming~\cite{reviewvanhecke}.  One prominent
signature of jamming is the singular behavior of the average number of
contacts per particle $z-z_J\propto (\phi - \phi_J)^{0.5}$, where
$z_J=2d$, $d$ being the space
dimension~\cite{ohernprl2002,contact_behringer}.  The distribution of
the gaps between particles displays a delta function at zero and a
square root decay for increasing gaps, which is key to the singular
behavior of the average contact
number~\cite{silbert_pre_2002_PhysRevE.65.031304,Zamponi_2010_RevModPhys.82.789,JB_PhysRevLett.106.135702,Donev_PRE_2005_PhysRevE.71.011105}.

\begin{figure}[t!] 
\onefigure{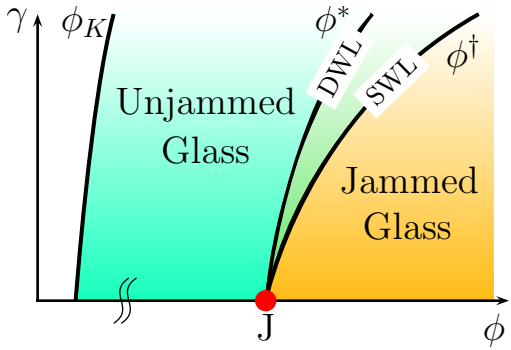}
\caption{{\bf Sketch of phase space.} (color online)
In the glass phase, two Widom lines, a structural one (SWL) and a
dynamical one (DWL) emerge from point J and separate the jammed
glass (frozen structure and frozen contacts) from the unjammed one (frozen
structure but "liquid" contacts).}
\label{fig:PhaseDiag}
\end{figure}

Although the average coordination number singularity is the hallmark
of jamming at zero temperature, its behavior is less clear at finite
temperature.  Both experimentally~\cite{zhang_vestiges2009} and
numerically,~\cite{zhang_vestiges2009,Xu_2011arXiv1112.2429W,JB_PhysRevLett.106.135702,Cheng_PRE_2010_PhysRevE.81.031301,hayakawa_JfiniteT_2011}
it has been observed that the first peak of the partial pair
correlation function has a finite maximum at a packing fraction
 $\phi_j(T)>\phi_j(0)=\phi_J$. This maximum has been
interpreted as a vestige of the divergence of the pair correlation
function at point J, the $T=0$ jamming transition. %(see
fig.~\ref{fig:PhaseDiag}).  However, it was later
argued~\cite{B926412D}, that this structural anomaly can be accounted
for using equilibrium liquid state theory, and is therefore not
specific to Jamming. The vicinity of point J has also been explored in
a mean-field-like replica description of thermal soft and hard
spheres~\cite{berthierjacquin_PRE}.  This description recovers all the
observed scalings in temperature and packing fraction \emph{but} the
square root singularity of the pair correlation function when $T=0^+$
and $\phi=\phi_J^+$. This discrepancy, together with the onset of a
diverging length in the vibrational properties of the jammed
state~\cite{wyart_EPL_2005_0295-5075-72-3-486}, suggest that larger
scale correlations must be taken into account, and calls for a better
characterization of the vicinity of point J.

% -- FIGURE 0000 ------------------------------

In the present letter, we focus on the \emph{dynamics} of the contact
network, a natural quantity of interest as soon as dynamics
  is present, which, to our knowledge, has never been explored so
far. To this end, we experimentally investigate a vibrated
two-dimensional bi-disperse packing of photo-elastic grains, close to
its jamming transition. We control both the packing fraction, $\phi$,
and the mechanical excitation, $\gamma$, to be defined precisely
below. This mechanically driven and dissipative system is
  far from equilibrium and the mechanical excitation is different from
  a temperature in many aspects. However, by analogy, one can reasonably
  consider that both temperature and mechanical excitation provide
  agitation to the particles. The structure, in terms of particle
neighborhoods, is completely frozen on the experimental time scales:
the system can safely be considered as a granular glass. The
structural signature of jamming $\phi^{\dagger}(\gamma)$ is given by
an abrupt variation of the average number of contacts.  Above
$\phi^{\dagger}(\gamma)$, the contacts are also frozen. On the
contrary below $\phi^{\dagger}(\gamma)$, they exhibit a rich dynamics
with strong heterogeneities.  For any finite mechanical excitation
$\gamma$, these heterogeneities are maximum at a packing fraction
$\phi^*(\gamma)$ smaller than $\phi^{\dagger}(\gamma)$. The relative
shift $|\epsilon^*|=|\phi^*-\phi^{\dagger}|/\phi^{\dagger}$ decreases
linearly towards zero and the characteristic length of the dynamical
heterogeneities eventually diverges as the mechanical excitation
approaches zero. Our results suggest the phase space diagram sketched
in Fig.~\ref{fig:PhaseDiag}, where the structural and the dynamical
crossovers are reminiscent of Widom lines~\cite{Stanley,Brazhkin} for
point J, namely the locus of the maxima of the second derivatives of
the free energy, introduced in the context of critical phenomena.

\begin{figure}[b!] 

\onefigure[width=0.9\columnwidth]{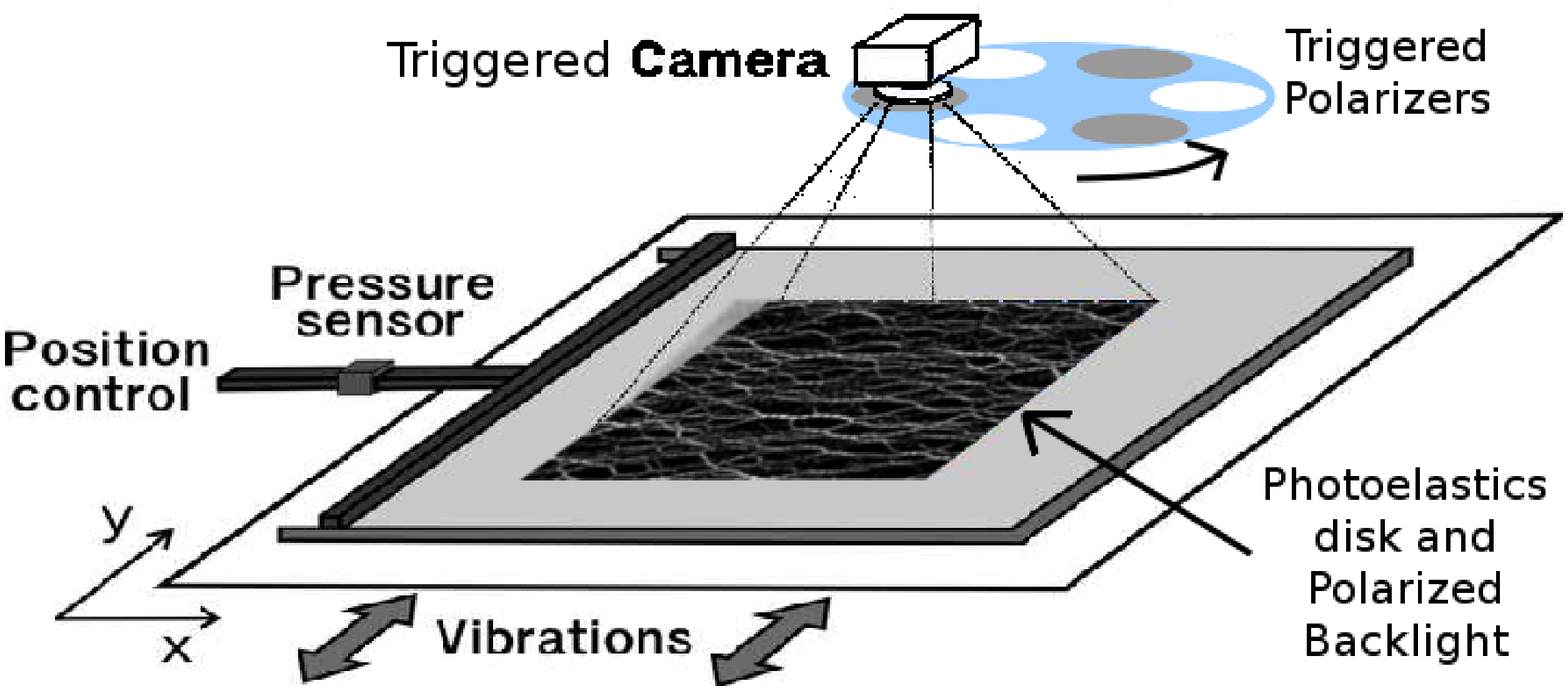}

\hspace{0.05\columnwidth}(a)

\mytwofigures[width=0.49\columnwidth,height=0.49\columnwidth]{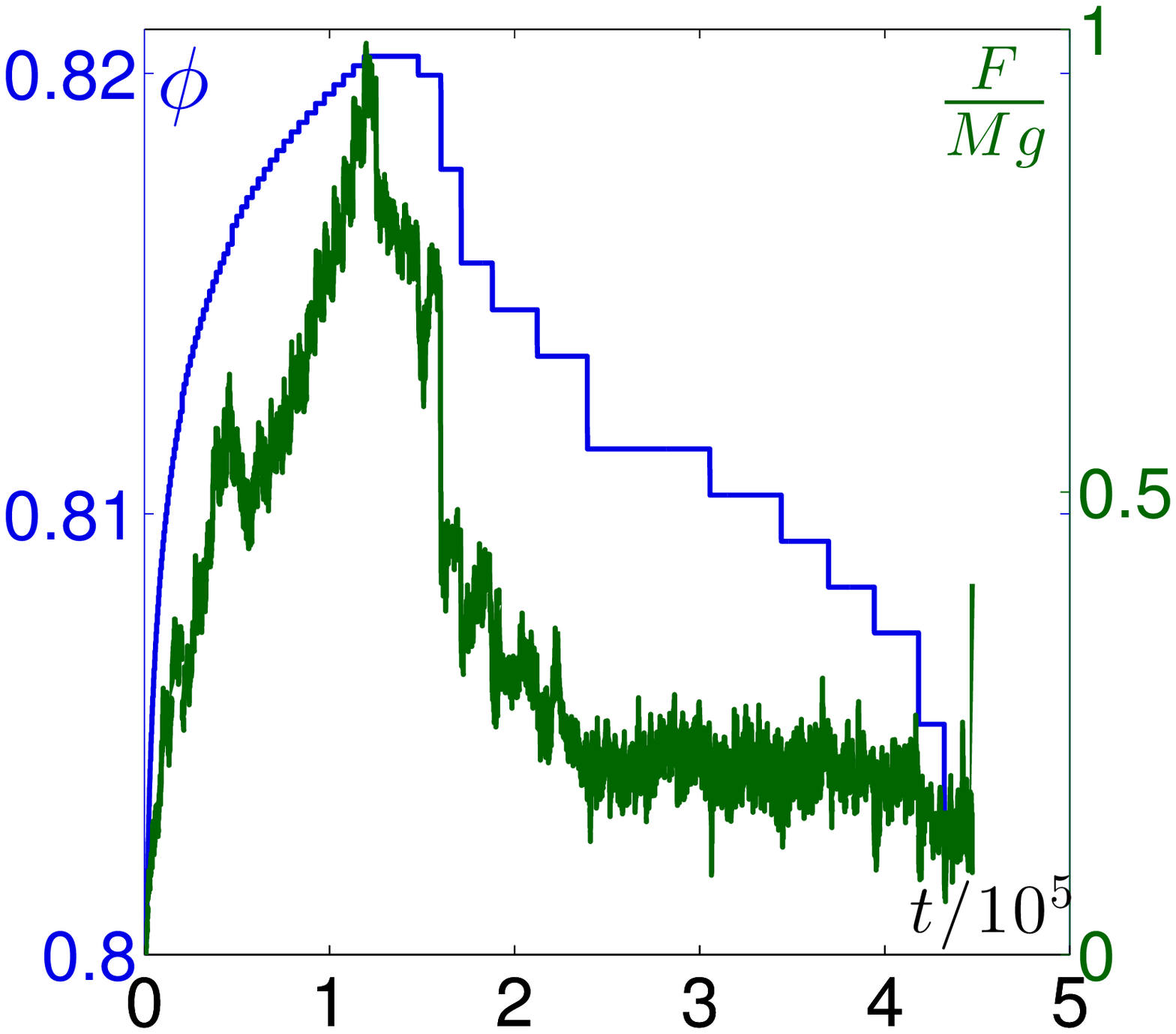}{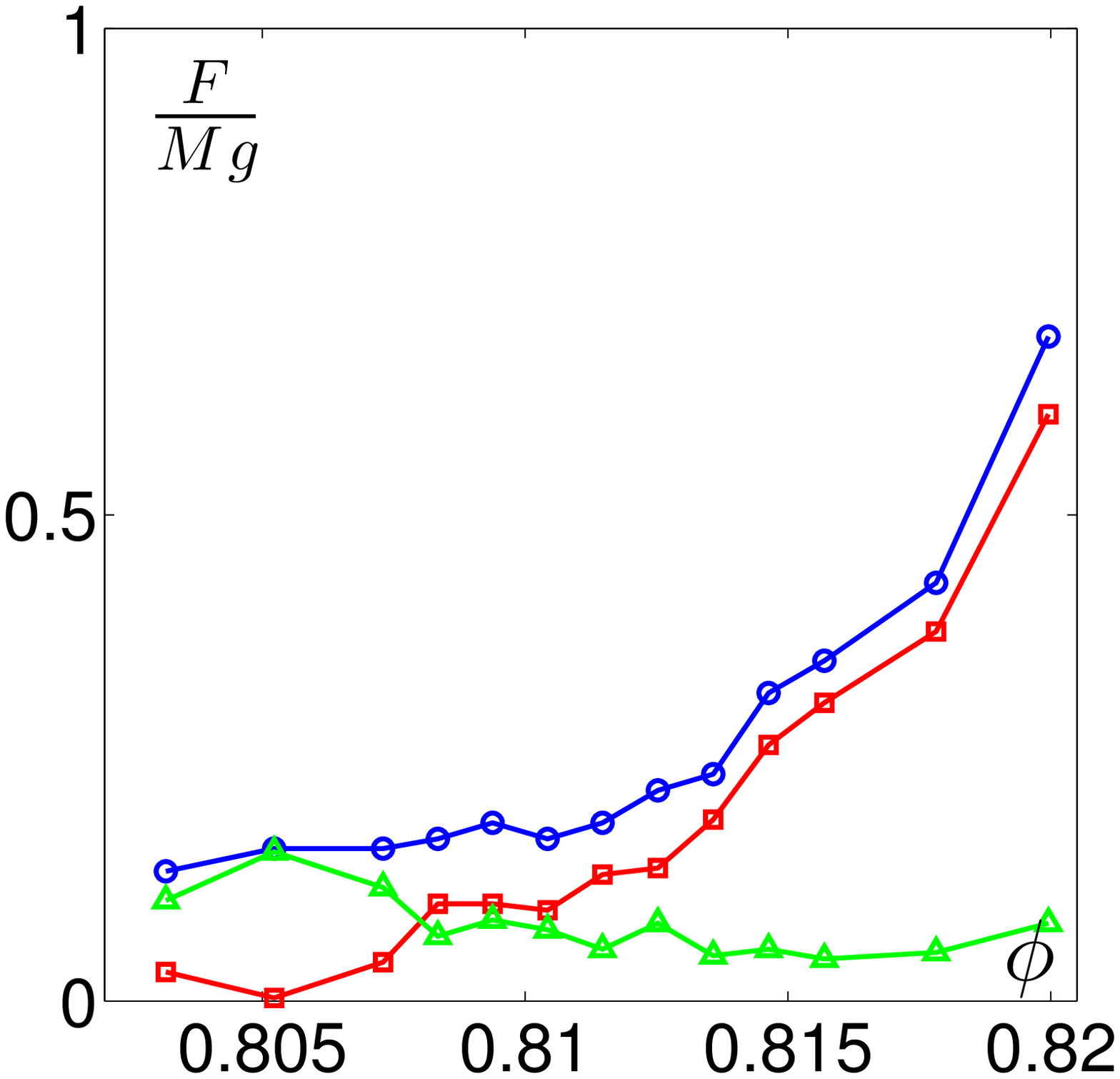}

\hspace{0.05\columnwidth}(b)\hspace{0.45\columnwidth}(c)

\caption{{\bf Experimental setup and protocol.} (color online) \leg{(a):} The
vibrating cell. \leg{(b):} Logarithmic increase followed by a stepwise decrease of the
packing fraction, and pressure at the wall.  \leg{(c)}: Average pressure vs
packing fraction, $\gamma=1.4$: (\textcolor{blue}{$\bigcirc$}) : $P_{TOT}$, (\textcolor{red}{$\square$}): $P_{STAT}$,
 (\textcolor{green}{$\triangle$}): $P_{DYN}$ as defined in the text.
  }
\label{fig:setup}
\end{figure}

Our experimental set up (Fig.~\ref{fig:setup} a) is adapted
from~\cite{lechenault_epl1,lechenault_epl2}, and allows us to shake
photoelastic discs between two cross-polarizers, thereby accessing the
contact and force
network~\cite{Majmudar_nature_2005,contact_behringer}.  A bi-disperse
mixture of~$4964$ and~ $3216$ polyurethane (PSM-4) discs, of
respectively ~$4 {\rm mm}$ and $5 {\rm mm}$~diameter, lies on a flat
transparent surface, which oscillates with an amplitude of~$1$~cm at
different frequencies of $6.25$, $7.5$ or $10$~Hz.  For frequencies
smaller than $f_0 =4.17$~Hz, the grains do not slip on the plate and
the mechanical excitation is effectively null.  
%This correspond to a grain-glass static friction of $0.7$. 
In the following we shall use the reduced frequency
$\gamma = (f - f_0)/f_0$
as a measure of the mechanical excitation. The grains are confined in
a static rectangular cell, for which the position of one horizontal
boundary is tuned using a micro-metric piston attached to a force
sensor.  A LED back-light and a circular polarizing sheet are inserted
in the vibrating surface, so that the grains are lighted by
transmission of circularly polarized light. A
monochromatic high resolution CCD camera records two types of images
in phase with the vibration. Every odd period, a circular analyzer is
inserted by a rotating wheel in the camera field to visualize the
photoelastic pattern.  From the direct-light images acquired every
even period, we obtain the grain positions with a resolution of
$0.5\%$ of the mean grain diameter. We then use standard tracking and
tessellation techniques to obtain the dynamics and structure of the
packings. From the cross-polarized images, we estimate the normal
force between two neighbors by integrating the square spatial gradient
of the light intensity over the area defined by the two Delaunay
triangles sharing a common edge.  Henceforth, all lengths are
expressed in units of the small grain diameter, and time is expressed
in units of vibration cycles.

% -- FIGURE 0002 ------------------------------

\begin{figure}[t!] 
\mytwofigures[width=0.47\columnwidth,height=0.47\columnwidth]{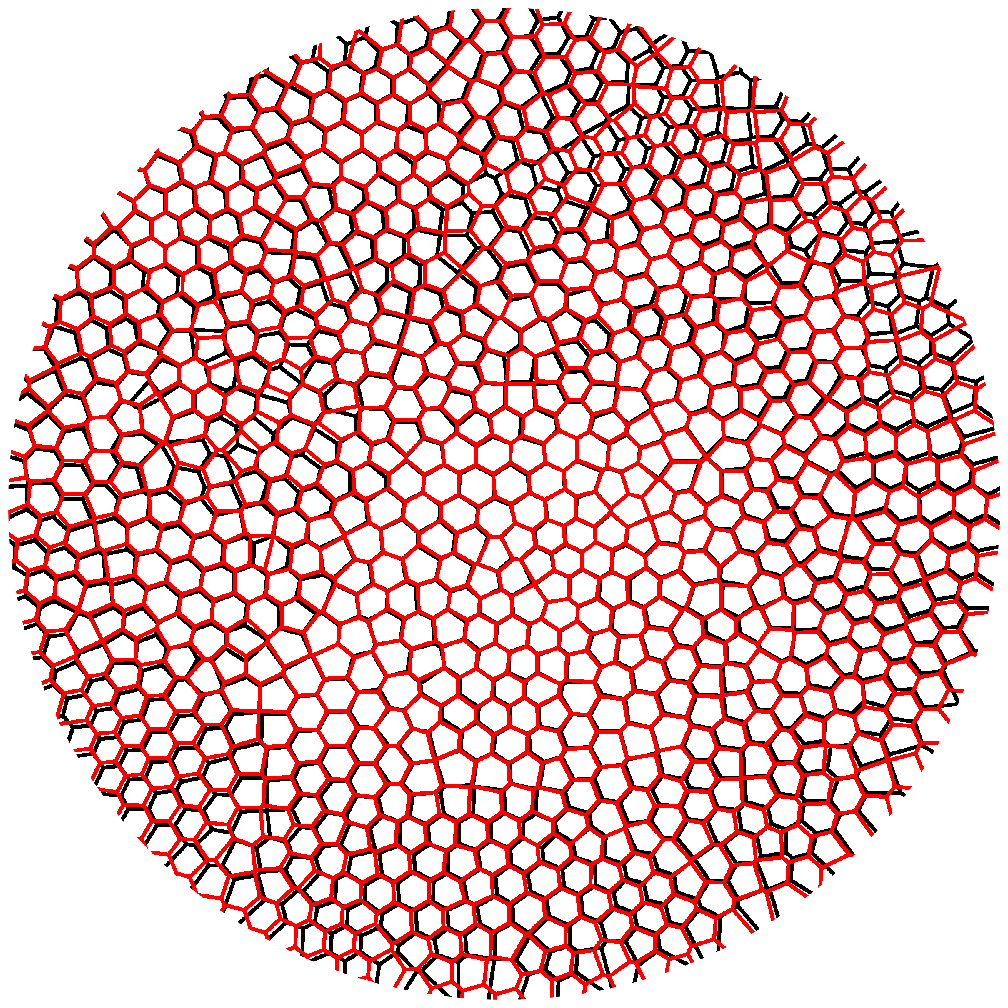}{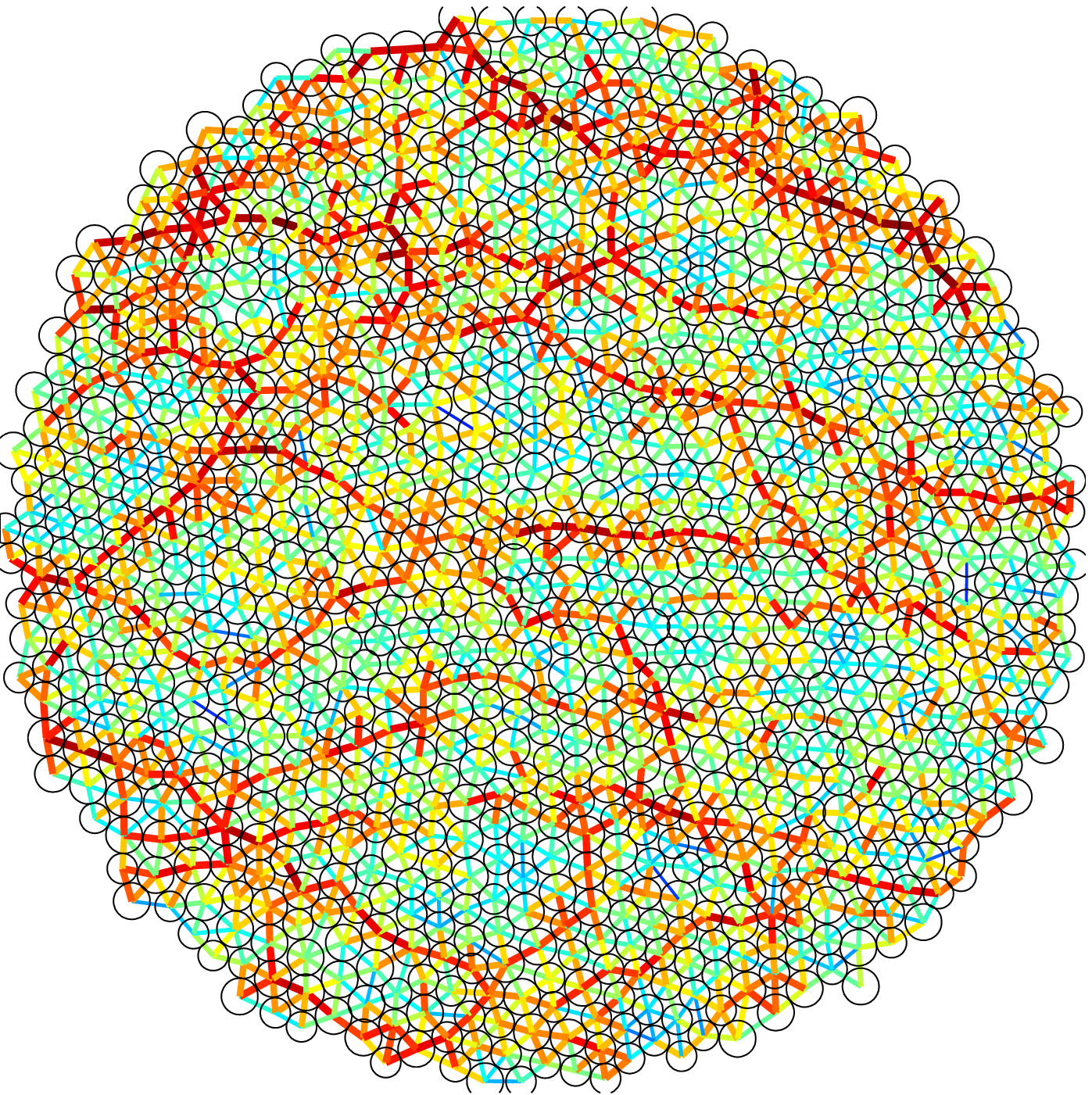}

\hspace{0.05\columnwidth}(a)\hspace{0.45\columnwidth}(c)

\mytwofigures[width=0.47\columnwidth,height=0.47\columnwidth]{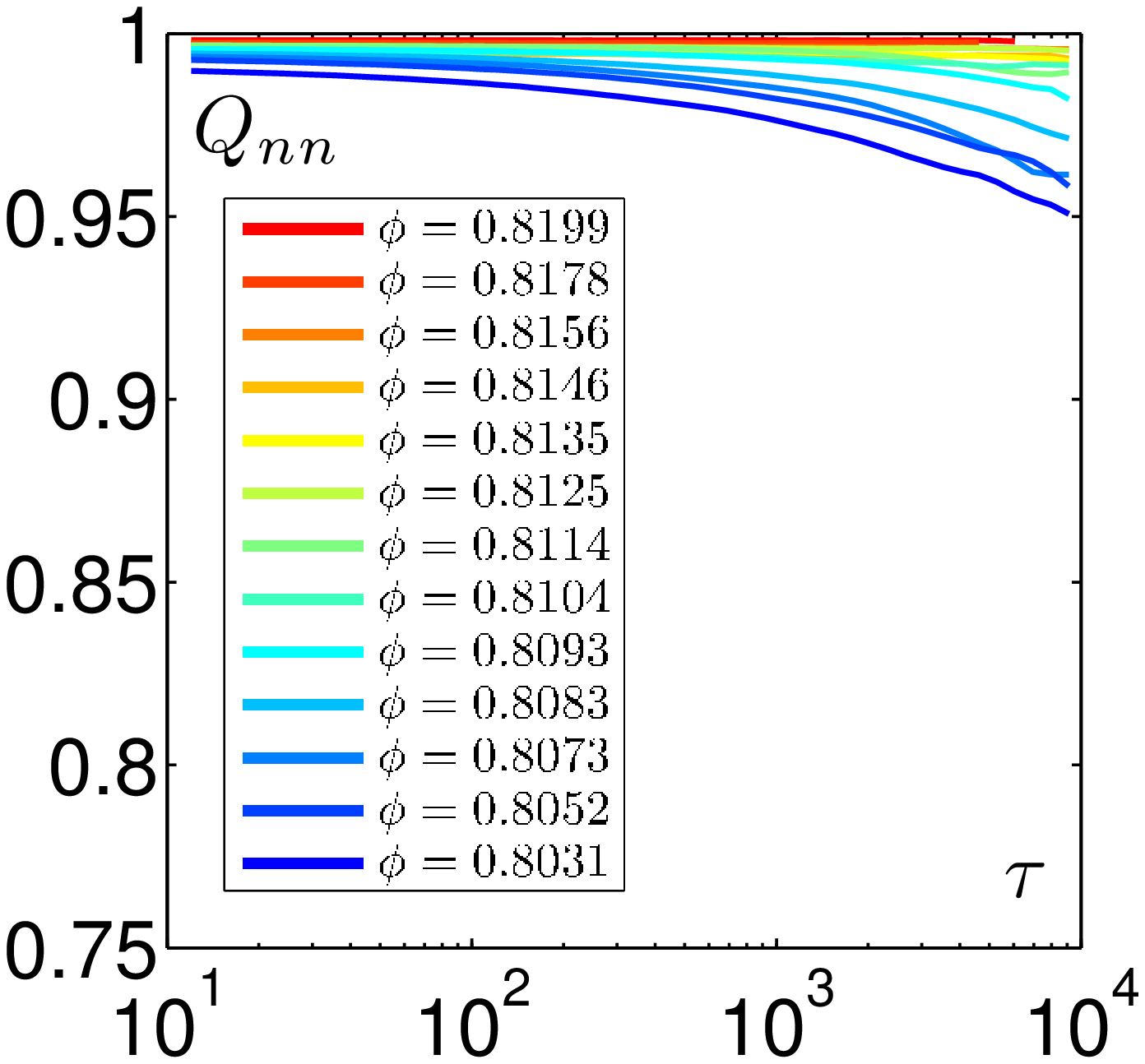}{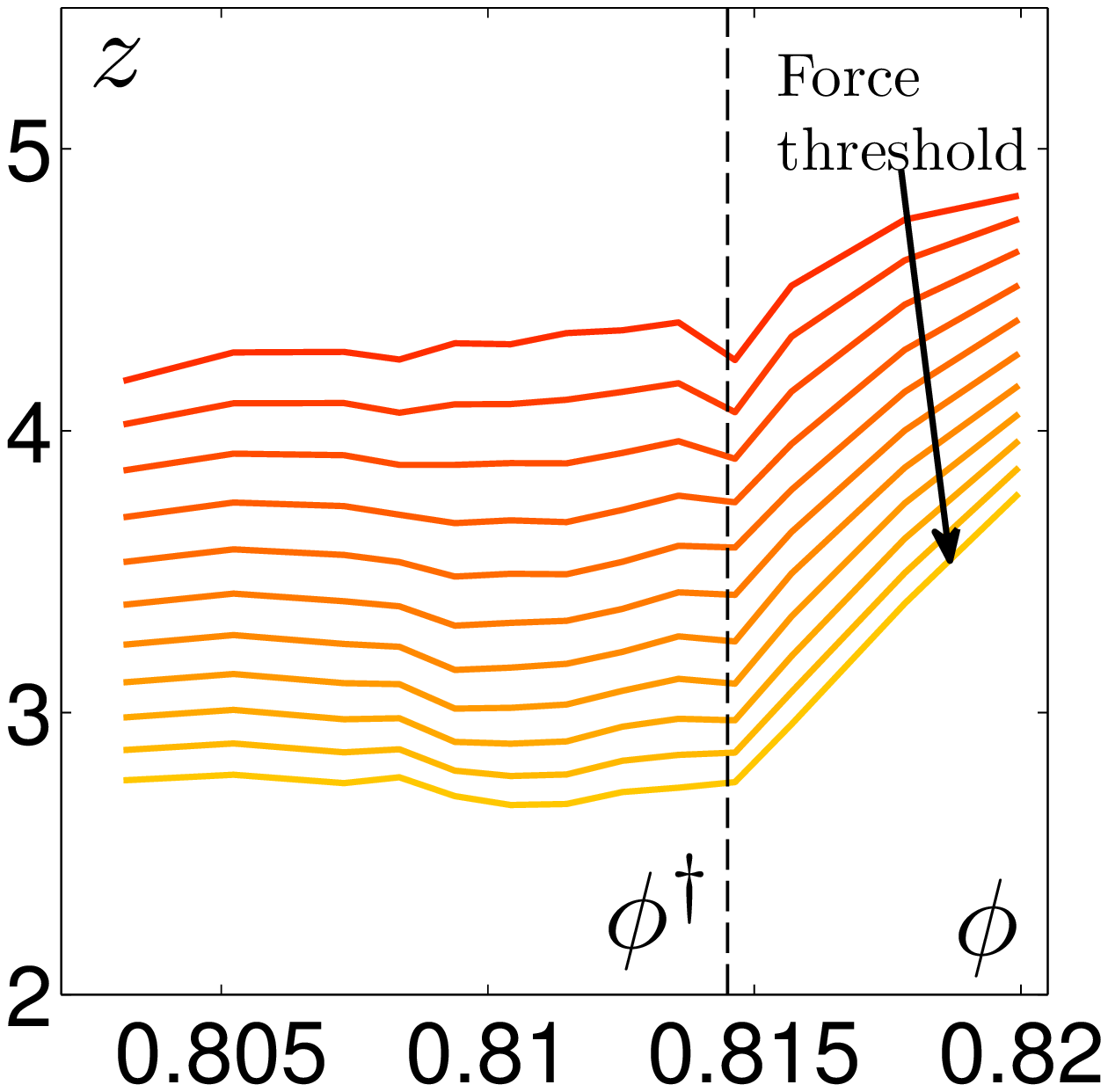}

\hspace{0.05\columnwidth}(b)\hspace{0.45\columnwidth}(d)

\caption{{\bf Structure of the granular glass, $\gamma=1.4$} (color online)
\leg{(a): } Superposition of the Laguerre cells computed at times
  $t=1$ and $t=5000$ for the loosest packing  ($\phi=0.8031$).   
\leg{(b): } Average fraction of neighbors  $Q_{nn}(\tau)$ which have not changed
  between two images separated by a time interval $\tau$, for different packing fractions, as
  indicated in the legend. 
\leg{(c) :} map of contact forces. Red colors stand for high forces and bluer color stand for low forces. 
\leg{(d) :}  Average coordination
  number $z$ vs. packing fraction $\phi$, determined according to
  various force thresholds }
\label{fig:glass}
\end{figure}

In order to ensure the highest and most reproducible jammed packings,
the packing fraction, which we control with a relative resolution of
$5 \times 10^{-6}$ is increased by steps of $\delta\phi =
3\times10^{-4}$ to some maximum value, using exponentially increasing
time steps (Fig.~\ref{fig:glass} b).  All images are then acquired
during stepwise decompression. For each decompression step we
carefully check that the system reaches a steady state, and all data
presented here have been computed in this steady
regime. Fig.~\ref{fig:glass} (c) displays the pressure with respect to
the packing fraction.  $P_{TOT}$ (respectively $P_{STAT}$) is the
pressure measured when the vibration is switched on (respectively
off).  $P_{STAT}$ is thus the static pressure sustained by the packing
in the absence of vibration, whereas $P_{DYN} = P_{TOT} - P_{STAT}$
can be interpreted as the additional dynamic pressure, induced by the
vibration.  At high packing fraction, the pressure is dominated by its
static part, whereas at low packing fraction, it is mostly dynamic. A
crossover occurs at some intermediate packing fraction, which
in~\cite{lechenault_epl1} was shown to coincide with a maximum in the
heterogeneity of the dynamics, as probed at scales of the order of
$10^{-2}$ grain diameter.  Here, the transition is shifted to lower
values of $\phi$, a fact which we attribute to the higher
interparticle friction of the photoelastic
discs~\cite{reviewvanhecke}.  Also the transitional range is wider,
which is likely due to the relative softness of the photoelastic discs
as compared to the brass discs used in earlier experiments.
(Young modulus of $0.5 {\rm GPa}$, as compared to $100 {\rm GPa}$).

% -- FIGURE 0003 ------------------------------

We first focus on the largest mechanical excitation level,
$\gamma=1.4$.  Altogether, the above compression protocol produces a
structure in terms of nearest neighbor relationships, which is
completely frozen on experimental timescales. The Laguerre
tessellation of two different packings separated by a time lag $5000$
can be superimposed almost perfectly, even at the loosest packing
fraction (Fig.~\ref{fig:glass} (a)). This is further quantified by
$Q_{nn}(\tau)$, the average fraction of \emph{neighbor} relationships
surviving in a time interval $\tau$. $Q_{nn}$ remains larger than
$95\%$ even for the loosest packing fraction, and is barely less than $100\%$
for the denser ones (Fig.~\ref{fig:glass}(b)). 
The contact network within this frozen structure is provided by
an analysis of the photoelastic images. A threshold is applied to both
the gap between neighboring particles and the contact force (see map
of contact forces in fig.~\ref{fig:glass} (c)), to decide what
particles are in contact.  Fig.~\ref{fig:glass} (d) displays the
average number of contacts, $z$, as a function of the packing fraction
for different thresholds. While the absolute value of $z$ depends on
the threshold, the dependence on $\phi$ is very robust: $z$ is
constant at low packing fractions, displays a kink at some
intermediate packing fraction, and thereafter, increases. The kink
occurs at a packing fraction which is independent of the threshold. We
interpret this packing fraction as the signature of a structural
crossover to jamming in the presence of vibration and find
$\phi^{\dagger}=0.8143 \pm 0.0005$.  Above $\phi^{\dagger}$, the
system is jammed and, $z\sim a\,(\phi -\phi^{\dagger})^\beta + z_c$,
where $\beta\in[0.4, 1]$ and $z_c\in[2.7, 4.3]$ depend on the
threshold.  Since counting arguments for frictional
packings~\cite{reviewvanhecke} constrain $z_c$ between $3$ and $4$ for
two-dimensional systems, it is fair to say that the transition
indicated by the kink is robust to the thresholding procedure. We now
come to the innovative part of this work, which consists of studying
the \emph{dynamics} of the contact network and its dependence on the
mechanical excitation.

 % -- FIGURE 0004 ------------------------------
 \begin{figure}[t!] 
\mytwofigures[height=0.48\columnwidth,height=0.48\columnwidth]{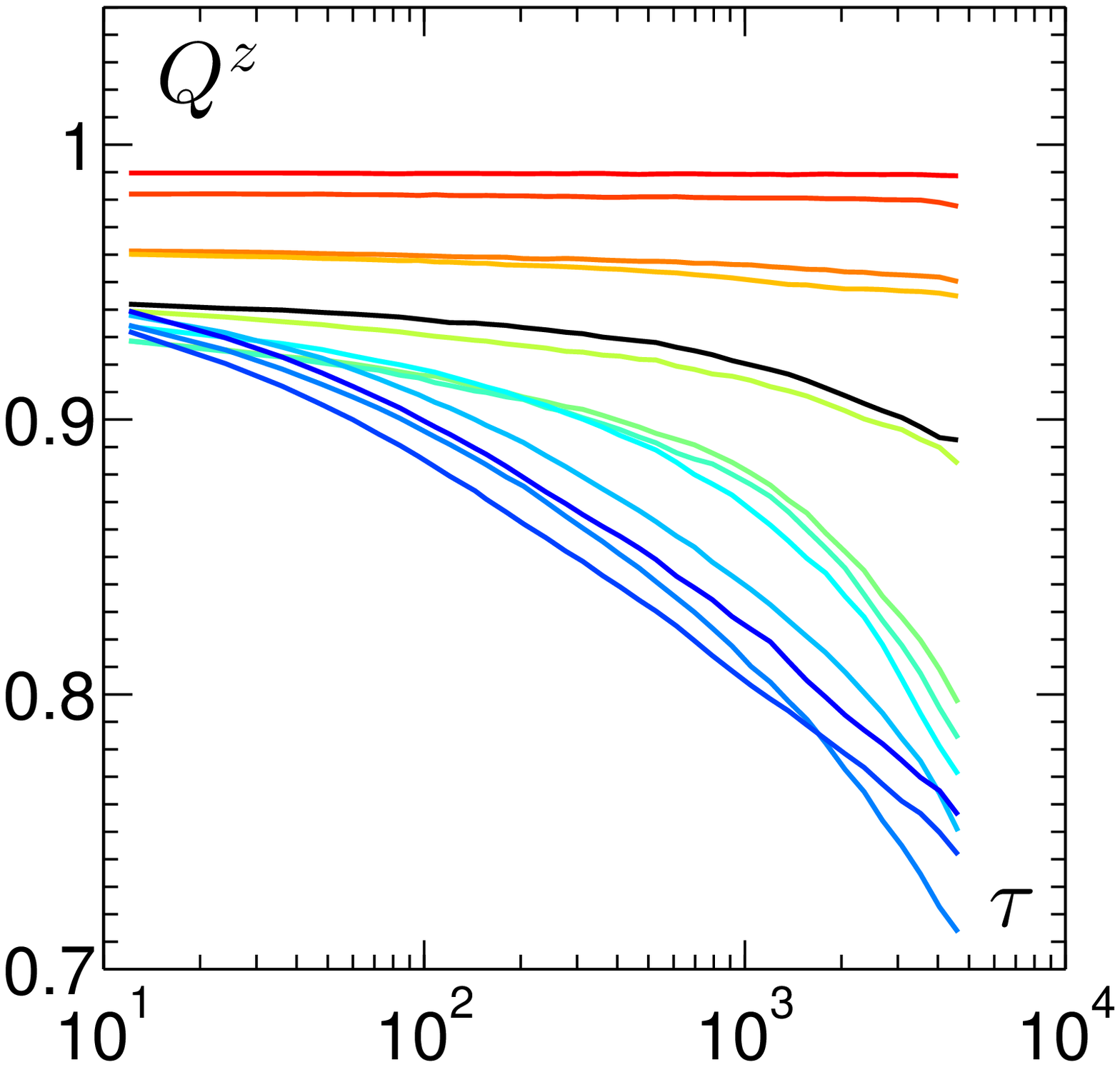}{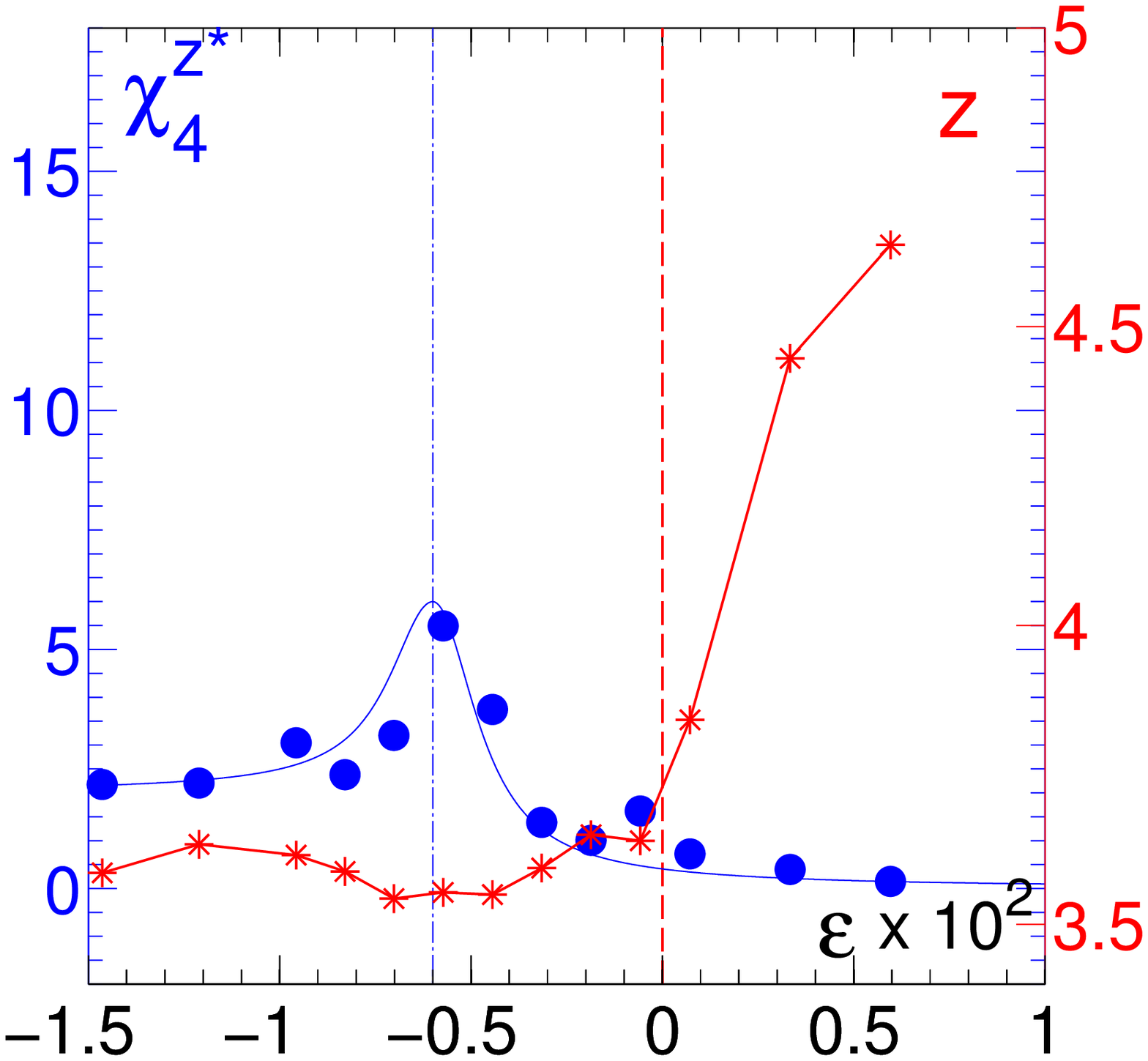}

\hspace{0.05\columnwidth}(a)\hspace{0.45\columnwidth}(d)

\mytwofigures[width=0.48\columnwidth,height=0.48\columnwidth]{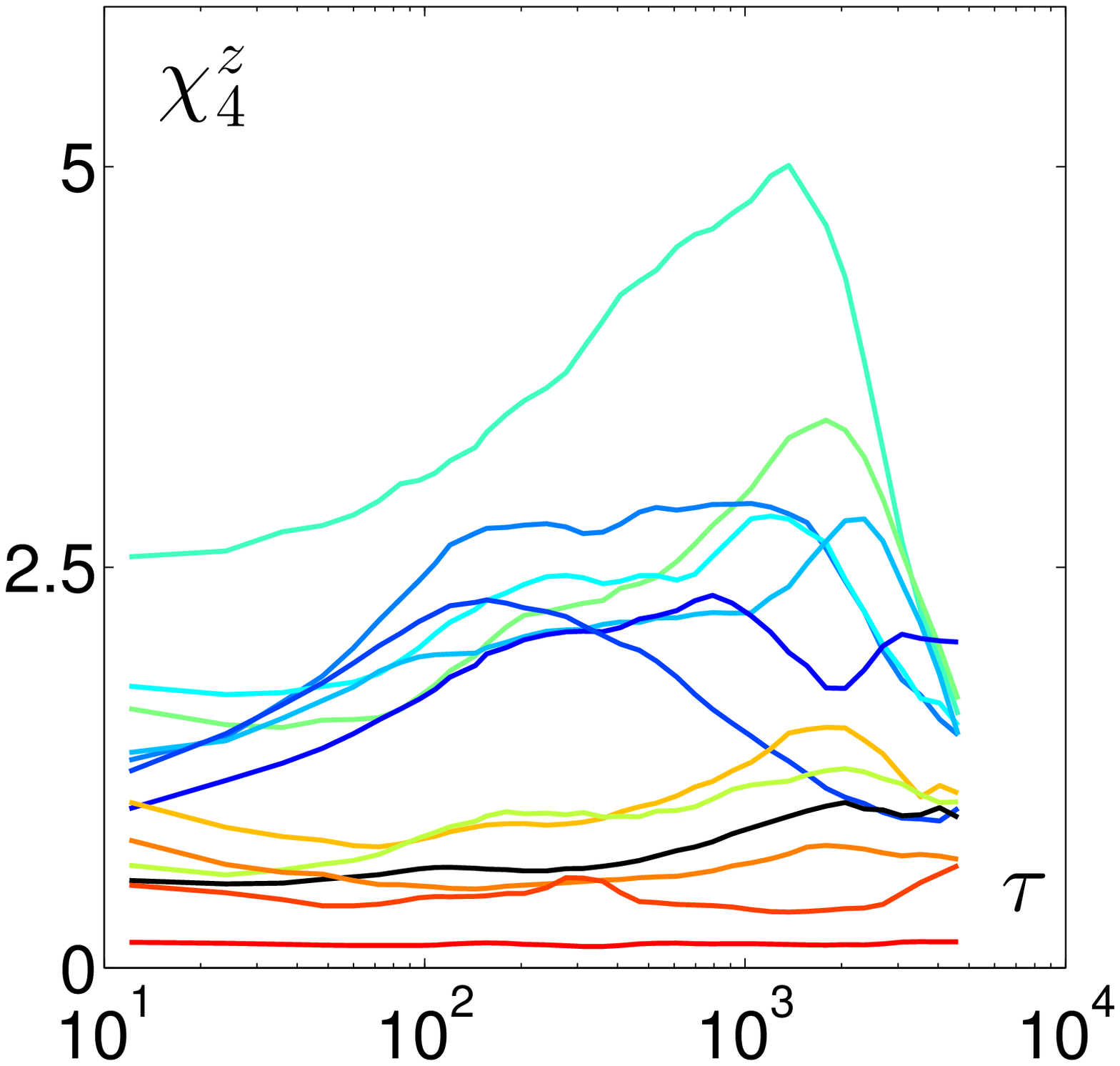}{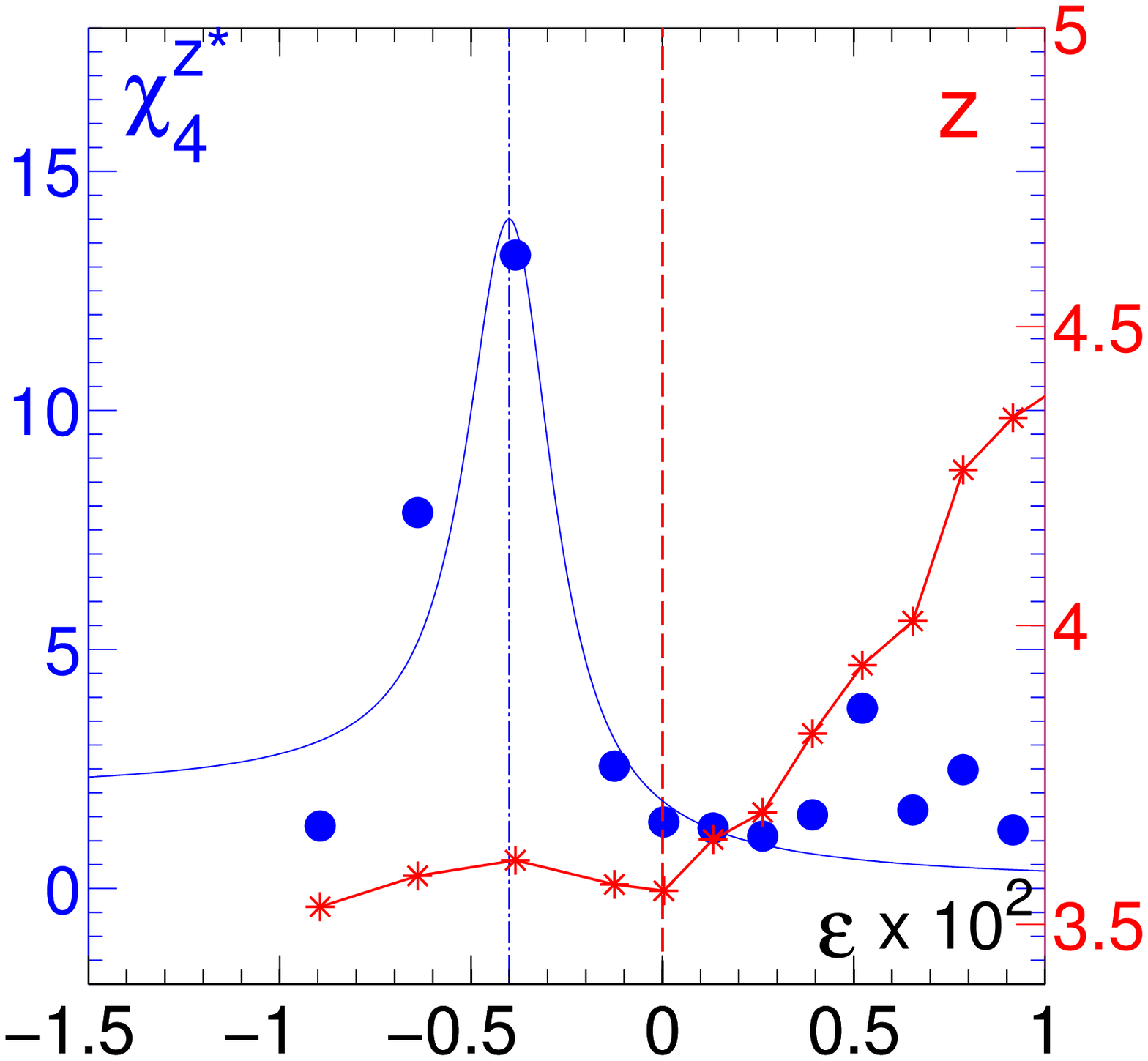}

\hspace{0.05\columnwidth}(b)\hspace{0.45\columnwidth}(e)

\mytwofigures[height=0.48\columnwidth,height=0.48\columnwidth]{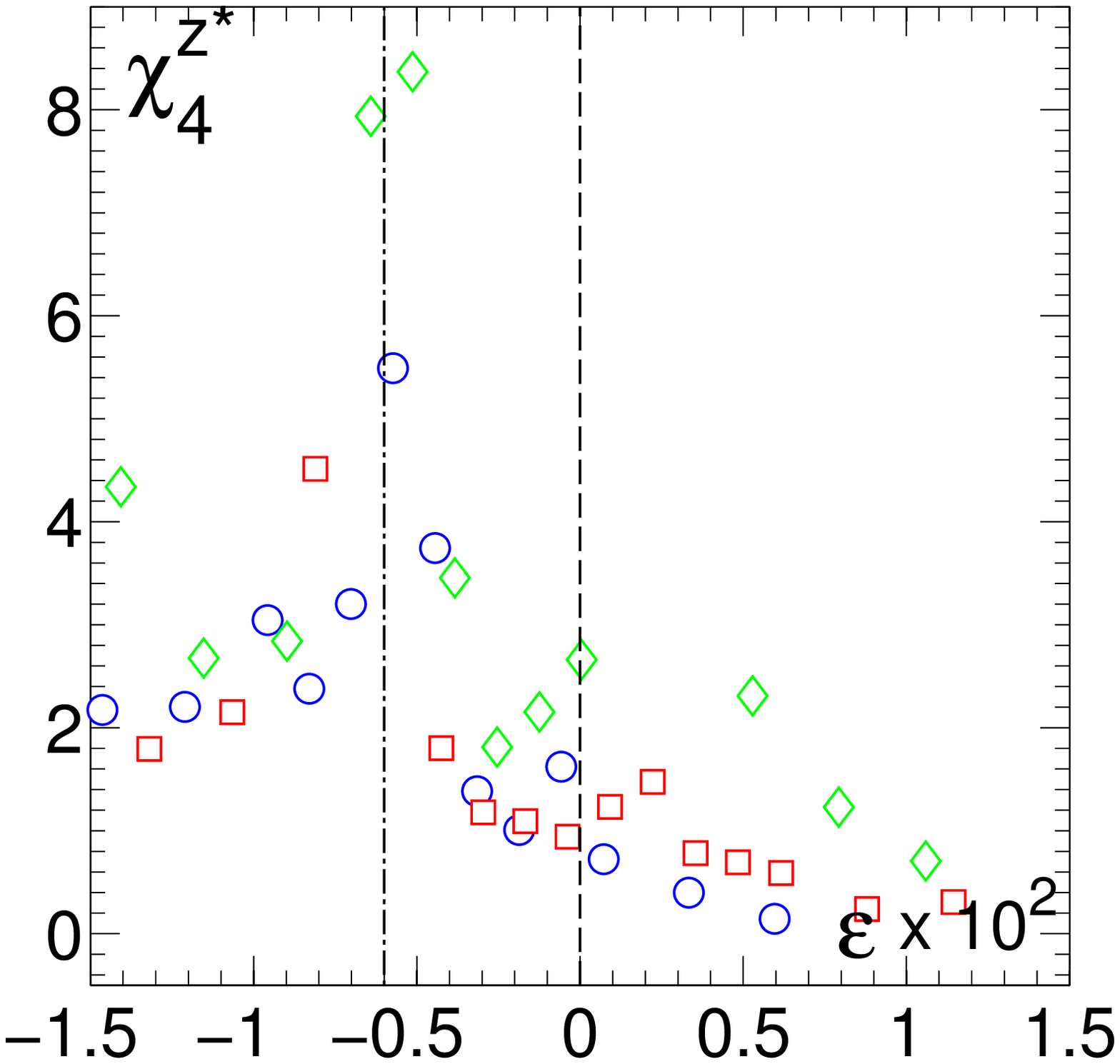}{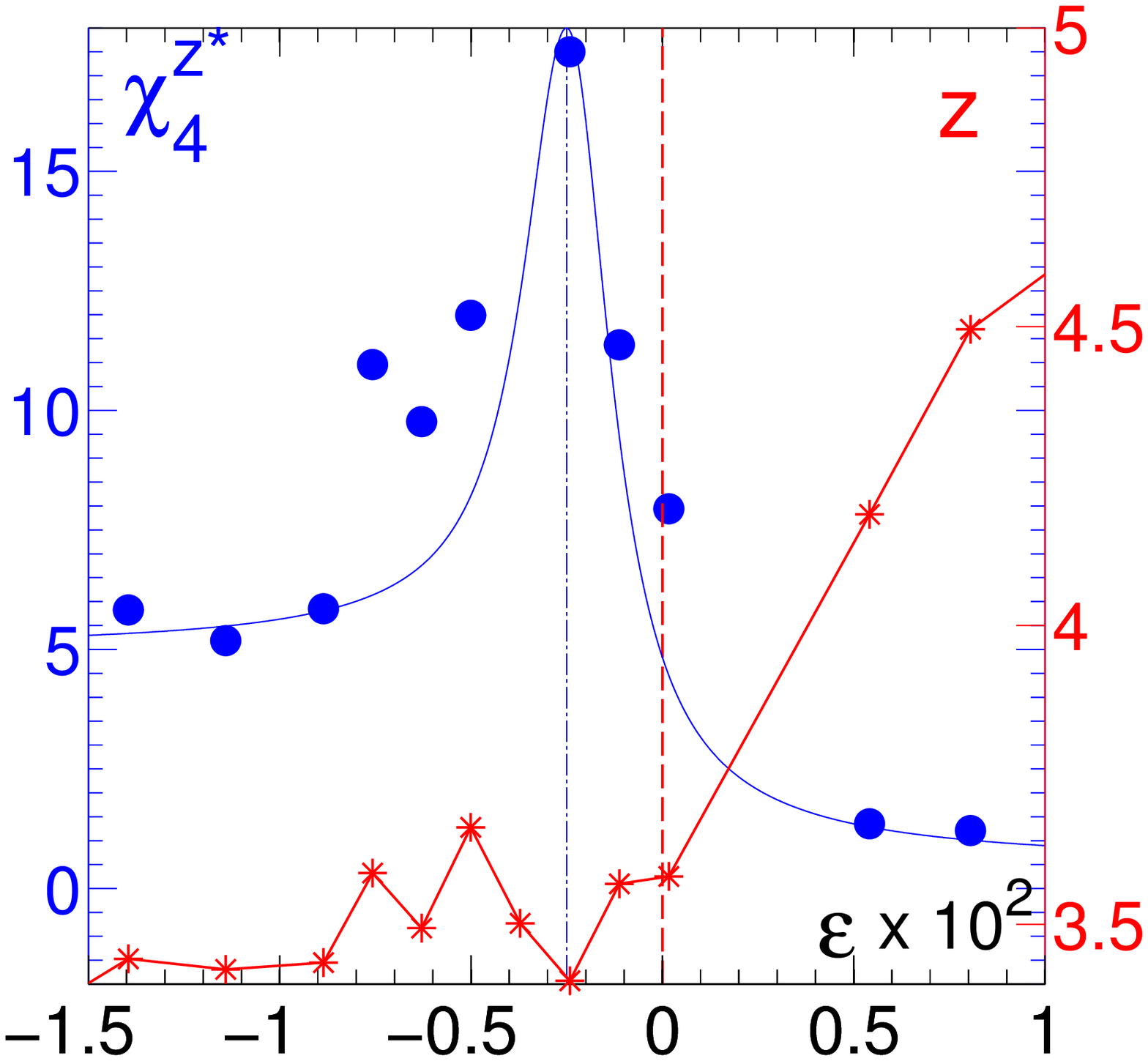}

\hspace{0.05\columnwidth}(c)\hspace{0.45\columnwidth}(f)

\caption{{\bf Contact Dynamics} (color online).  \leg{(a):} Average
  contact overlap $Q_{z}(\tau)$ for different packing fractions,
  $\gamma=1.4$ (same color code as in fig.~\ref{fig:glass} (a) with
  the curve corresponding to $\phi_J$ plotted in black).  \leg{(b):}
  Contact overlap dynamical susceptibility $\chi_4^{z}(\tau)$,
  $\gamma=1.4$ (same color code).  \leg{(c):} Contact overlap
  dynamical susceptibility ${\chi_4^z}^*$, chosen for the delay time, $\tau$, that maximizes
  its value, as a function of $\epsilon=(\phi^*-\phi^{\dagger})/\phi^{\dagger}$ for three realizations
  $(\textcolor{blue}{\bullet},\textcolor{red}{\square},\textcolor{green}{\diamond})$
  of the same experiment, $\gamma=1.4$.  \leg{(d,e,f):} Maximal
  contact overlap dynamical susceptibility ${\chi_4^z}^*
  (\textcolor{blue}{\bullet})$ and average contact number $z
  (\textcolor{red}{\star})$ versus $\epsilon$ for three different
  amplitudes of the vibration: (d) $\gamma=1.4$; (e) $\gamma=0.8$; (f)
  $\gamma=0.5$).  The red, respectively blue, dashed line indicates the
  location of $\phi^*$, resp. $\phi^{\dagger}$.  The continuous blue curve is
  a guide to the eye. }
\label{fig:dynamics}
\end{figure}

Analyzing the force network, we recover existing results on the force
distributions\cite{ohernPRL01} and observe that its dynamics is slaved
to that of the contacts.  We thus concentrate on the description of
the dynamics of the contact network.  It is naturally quantified by an
estimator of the contact overlap between $t$ and $t+\tau$: \be
Q^{z}(t,\tau)=\frac{1}{N}\sum_i Q^{z}_{i}(t,\tau), \ee where
$Q^{z}_{i}(t,\tau)=\Theta(2-|\delta z_i(t,\tau)|)$, with $\Theta(.)$,
the Heavyside function and $\delta z_i(t,\tau)$, the change in number
of contact of grain $i$, between $t$ and $t+\tau$.  Alternative
definitions, e.g. requiring smaller or larger local contact
fluctuations, do not change the following conclusions.
Fig.~\ref{fig:dynamics} (a) displays $Q_{z}(\tau)=\langle
Q^z(t,\tau)\rangle_t$ for the various packing fractions, where
$\langle . \rangle_t$ denotes the time average. The black curve
corresponds to the packing fraction of the jamming crossover
$\phi^{\dagger}$.  For $\phi>\phi^{\dagger}$, $Q^z(\tau)$ remains
constant at values ranging between $0.7$ and $0.9$, indicating that
there is no long-time decorrelation of the contact network: the
contacts are established permanently, once they are formed.  The sole
decorrelation observed above $\phi^{\dagger}$ occurs at short times
and is induced by the fast dynamics of the rattlers, the number of
which increases when $\phi$ approaches $\phi^{\dagger}$.  For
$\phi<\phi^{\dagger}$, long time relaxation clearly sets in,
indicating that contacts now rearrange. On can thus think about this
structural crossover as a glass transition for the binary degrees of
freedom (e.g. yes/no) which indicate whether neighboring particles are
in contact or not.  Accordingly, one would like to further investigate
this long term dynamics, in order to see whether it shares other
similarities with glassy dynamics in spin systems.

We compute the dynamical susceptibility (see~\cite{DH-book} for an introduction
to dynamical heterogeneities): 
\be
\chi_4^{z}(\tau) = N \frac{Var(Q^{z}(t,\tau))}{\langle
  Var(Q^{z}_{i}(t,\tau))\rangle_i},
\label{eq:Chi}
\ee 
where $Var(.)$ denotes the variances sampled over time and $\langle. \rangle_i$
denotes the average over the grains. This dynamic susceptibility estimates
the range of the spatial correlation in the dynamics of the contact network.
One sees in Fig.~\ref{fig:dynamics} (b) that $\chi_4^{z}(\tau)$ becomes
significant for $\phi<\phi^{\dagger}$ and then exhibits a maximum ${\chi_4^z}^*$
in time, which in turn displays a clear maximum at a packing fraction $\phi^*$.
Performing three independent experimental runs, with the same vibration $\gamma=1.4$, 
we observe in fig.~\ref{fig:dynamics}(c) that  $\phi^*$ is systematically
smaller than $\phi^{\dagger}$, with a relative shift $\epsilon^*=(\phi^*-\phi^{\dagger})/\phi^{\dagger} = -5\times10^{-3}$ : the reorganization of the contacts is maximally collective, indicating
a dynamical crossover \emph{below} $\phi^{\dagger}$. Since it is well known that the jamming
transition and its related crossovers, a priori depend on the initial conditions and the preparation protocol~\cite{PhysRevLett.104.165701}, we shall now use the reduced packing fraction $\epsilon=\phi-\phi^{\dagger})/\phi^{\dagger}$, in order to compare different experimental runs.

% -- FIGURE 0005 ------------------------------
\begin{figure}[t!] 
\mytwofigures[width=0.48\columnwidth,height=0.48\columnwidth]{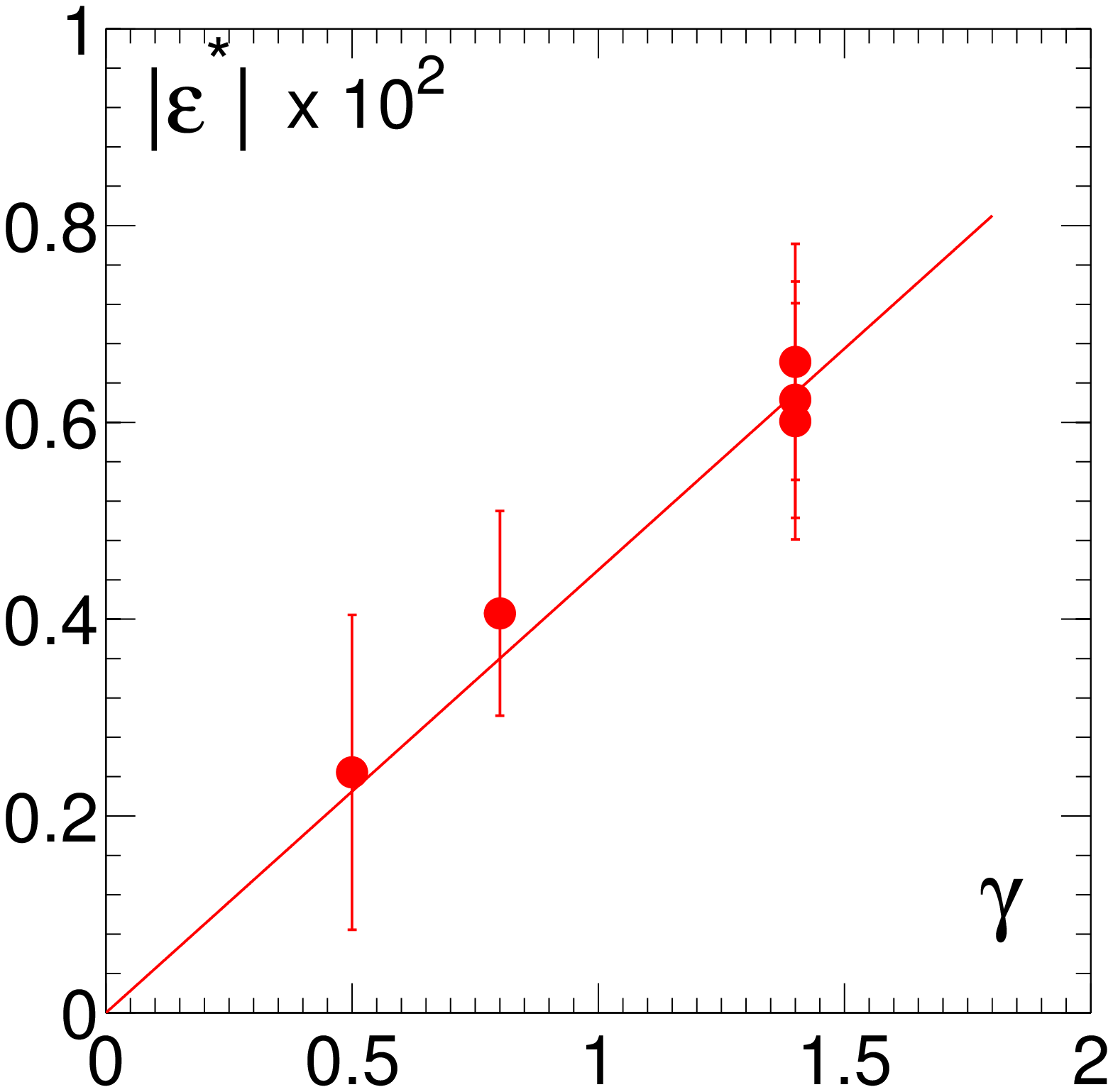}{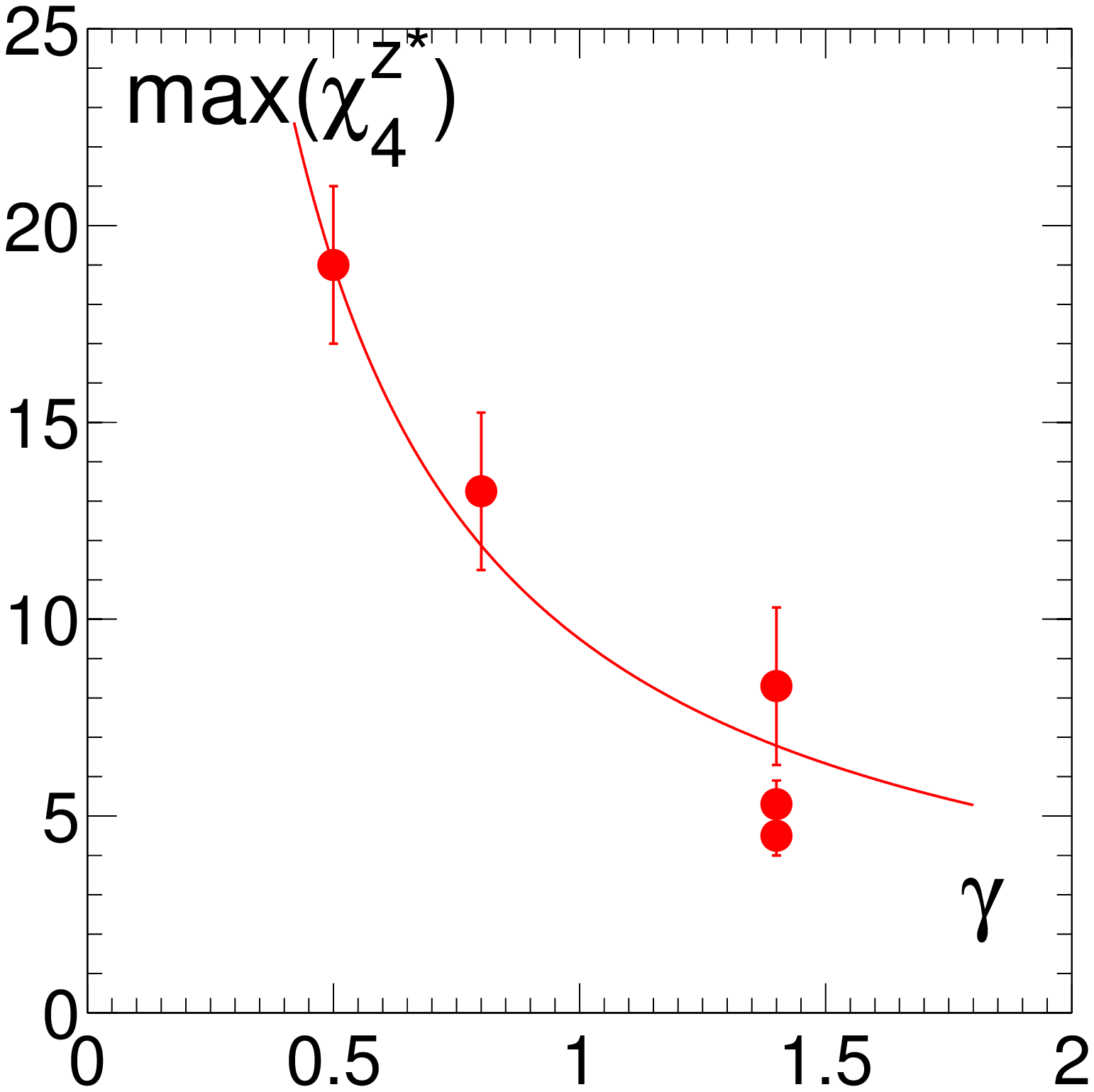}

\hspace{0.05\columnwidth}(a)\hspace{0.45\columnwidth}(b)

\caption{{\bf Dependence on the vibration amplitude} (color online) of
  \leg{(a)} $|\epsilon^*|$, the relative shift between $\phi^*$ and
  $\phi^{\dagger}$ and \leg{(b)} the maximal dynamical susceptibility as a
  function of the vibration amplitude $\gamma$. The continuous lines
  indicate, respectively, a linear and a $1/\gamma$ dependence on
  $\gamma$.}
\label{fig:vibration}
\end{figure}

The robustness of our finding is further reinforced by the fact that
we could demonstrate a systematic dependence on the excitation
$\gamma$.  Figures~\ref{fig:dynamics}(d,e,d) display the dynamical
susceptibility together with the average contact number as a function
of the packing fraction for respectively $\gamma=1.4, 0.8, 0.5$. On
decreasing $\gamma$, one observes that (i) the dynamical crossover
$\phi^*$ moves closer to the structural signature of jamming and (ii)
the maximum of the dynamical susceptibility ${\chi_4^z}^*$, hence the
correlation length, increases when $\gamma$
decreases. Figure~\ref{fig:vibration} summarizes these
dependancies. Our data are compatible with a linear variation
$|\epsilon^*| \propto \gamma$ and $\textrm{max}{(\chi_4^z}^*)\propto
1/\gamma$ suggesting a divergence of the correlation length at zero
excitation.

Let us now summarize and discuss our observations. Having prepared a
granular glass, with a frozen neighborhood structure, we have observed
three salient features of the contact network: (i) the evolution of
the averaged number of contacts with packing fraction, $z(\phi)$,
points to a first transitional packing fraction $\phi^{\dagger}$; (ii)
the dynamics of the contact network, together with the fluctuation of
the coordination number are maximally heterogeneous at a packing
fraction $\phi^*<\phi^{\dagger}$.  The shift in packing fraction
decreases linearly with the amplitude of the mechanical excitation,
while the dynamical heterogeneities increase sharply.  This
phenomenology, summarized in Fig.~\ref{fig:PhaseDiag}, is reminiscent
of the so-called Widom lines observed in the supercritical region
close to a critical point~\cite{Stanley,Brazhkin}.  To our knowledge,
the present study is the first experimental characterization of these
crossovers, including their dependence on the mechanical
excitation. Obviously one would like to extend the range of dependence
towards even lower excitation, but the required precision then calls
for numerical investigations.

Along the way, our results address a long standing conundrum left from
earlier experiments using the same apparatus, but with hard (brass)
discs~\cite{lechenault_epl1}.  The authors observed a maximum in the
heterogeneities of the dynamics for the packing fraction, where
$P_{DYN}(\phi)$ and $P_{STAT}(\phi)$ intersect.
% They correctly attributed their observation to jamming, but  they could not precisely
%identify the underlying mechanism responsible for these heterogeneities. Here, we were
%in the position to attribute them to the dynamics of the contact network.
The existence of this maximum suggests that the experiment probed both
sides of the jamming transition, a puzzling conclusion given the very
strong stiffness of the brass discs. Using soft discs, we demonstrate
here that there are several signatures of point J at finite mechanical
excitation, $\gamma$, and that the one associated with the dynamical
heterogeneities occurs at a lower packing
fraction, $\phi^*(\gamma)$, than the one at which the average number
of contact increases, $\phi^{\dagger}(\gamma)$.  The previous
experiment using brass discs~\cite{lechenault_epl1} was actually
probing the dynamical crossover, $\phi^*$, both sides of which lie
below the structural signature of the jamming transition.  This is
further confirmed by the observation that, here also, $P_{DYN} \simeq
P_{STAT}$ at $\phi^*$.

Unlike thermal systems, our system is in an
  out-of-equilibrium, mechanically driven state. Still, recent
  numerical simulations~\cite{Xu_2011arXiv1112.2429W,
    JB_PhysRevLett.106.135702, hayakawa_JfiniteT_2011, Ikeda:2012wp}
  suggest that for the kind of physics we are interested in, the
  similarities with thermal systems are much stronger than one may
  have expected at first sight. For instance, the structural crossover
  reported here might be related to the finite temperature first-peak
  pair-correlation maxima near the jamming point reported
  in~\cite{zhang_vestiges2009,JB_PhysRevLett.106.135702,Xu_2011arXiv1112.2429W}.
  More specifically, in~\cite{Ikeda:2012wp} the authors report an
  extensive study of the dynamics close to point J, in the
  temperature-density space, which they conclude by comparing with
  existing colloidal experiments. To do so they essentially use the
  Debye-Waller factor, namely the size of the cage surrounding the
  particles, as a sensitive thermometer. In the present case, the cage
  size at the largest packing fraction and largest magnitude of
  excitation is of the order of $5 \times 10^{-3}$. This would
  correspond in Fig. 1 of their work to a rescaled kinetic energy of
  $10^{-6}$.  For lower excitation, it is even smaller.  In all cases,
  we conclude that the present experiments, together with those
  of~\cite{lechenault_epl1,lechenault_epl2} are for the moment the
  only ones, which have probed the dynamical criticality related to the
  jamming transition (see the discussion part and Fig. 12
  in~\cite{Ikeda:2012wp}).  Whether the same scenario as the one
  described here holds for thermal soft spheres, such as emulsions, is
  an open issue for further experimental investigation.

Finally, one cannot exclude the possible effect of
  friction. Here the friction coefficient amongst the grains is
  typically $\mu=0.7$ and one indeed notes that the packing fractions
  of interest reported here have a lower value than those obtained for
  the brass discs ($\mu=0.4$) and for frictionless particles. However
  our results demonstrate that at the qualitative level and in the
  dynamical regime probed by our set-up, friction does not seem to be
  a relevant parameter. Whether it impacts the quantitative scaling
  properties close to points J requires further studies, presumably
  numerical ones.

\acknowledgements
We acknowledge L. Berthier and F. Zamponi for illuminating discussions and are grateful to C\'ecile Wiertel-Gasquet and Vincent Padilla for
their skillful technical assistance.

\bibliography{biblio_twotransitions}

\begin{thebibliography}{10}
\expandafter\ifx\csname url\endcsname\relax\def\url#1{\texttt{#1}}\fi

\bibitem{reviewvanhecke}
\Name{van Hecke M.} \REVIEW{Journal of Physics: Condensed
  Matter}{22}{2010}{033101}.

\bibitem{ohernprl2002}
\Name{O'Hern C.~S., Langer S.~A., Liu A.~J. \and Nagel. S.~R.} \REVIEW{Phys.
  Rev. Lett.}{88}{2002}{075507}.

\bibitem{contact_behringer}
\Name{Majmudar T.~S., Sperl M., Luding S. \and Behringer R.~P.} \REVIEW{Phys.
  Rev. Lett.}{98}{2007}{058001}.

\bibitem{silbert_pre_2002_PhysRevE.65.031304}
\Name{Silbert L.~E., Erta\ifmmode~\mbox{\c{s}}\else \c{s}\fi{} D., Grest G.~S.,
  Halsey T.~C. \and Levine D.} \REVIEW{Phys. Rev. E}{65}{2002}{031304}.

\bibitem{Zamponi_2010_RevModPhys.82.789}
\Name{Parisi G. \and Zamponi F.} \REVIEW{Rev. Mod. Phys.}{82}{2010}{789}.

\bibitem{JB_PhysRevLett.106.135702}
\Name{Jacquin H., Berthier L. \and Zamponi F.} \REVIEW{Phys. Rev.
  Lett.}{106}{2011}{135702}.

\bibitem{Donev_PRE_2005_PhysRevE.71.011105}
\Name{Donev A., Torquato S. \and Stillinger F.~H.} \REVIEW{Phys. Rev.
  E}{71}{2005}{011105}.

\bibitem{zhang_vestiges2009}
\Name{Zhang Z., Xu N., Chen D. T.~N., Yunker P., Alsayed A.~M., Aptowicz K.~B.,
  Habdas P., Liu A.~J., Nagel S.~R. \and Yodh A.~G.}
  \REVIEW{Nature}{459}{2009}{230}.

\bibitem{Xu_2011arXiv1112.2429W}
\Name{{Wang} L. \and {Xu} N.} \REVIEW{ArXiv e-prints}{}{2011}{}.

\bibitem{Cheng_PRE_2010_PhysRevE.81.031301}
\Name{Cheng X.} \REVIEW{Phys. Rev. E}{81}{2010}{031301}.

\bibitem{hayakawa_JfiniteT_2011}
\Name{Otsuki M. \and Hayakawa H.} \REVIEW{Phys. Rev. E}{86}{2012}{031505}.

\bibitem{B926412D}
\Name{Jacquin H. \and Berthier L.} \REVIEW{Soft Matter}{6}{2010}{2970}.

\bibitem{berthierjacquin_PRE}
\Name{Berthier L., Jacquin H. \and Zamponi F.} \REVIEW{Phys. Rev.
  E}{84}{2011}{051103}.

\bibitem{wyart_EPL_2005_0295-5075-72-3-486}
\Name{Wyart M., Nagel S.~R. \and Witten T.~A.} \REVIEW{EPL (Europhysics
  Letters)}{72}{2005}{486}.

\bibitem{Stanley}
\Name{Stanley H.~E.} \Book{Introduction to Phase Transitions and Critical
  Phenomena.} (Oxford University Press) 1971.

\bibitem{Brazhkin}
\Name{Brazhkin V.~V., Fomin Y.~D., Lyapin A.~G., Ryzhov V.~N. \and Tsiok E.~N.}
  \REVIEW{J. Phys. Chem. B}{115}{2011}{14112–14115}.

\bibitem{lechenault_epl1}
\Name{Lechenault F., Dauchot O., Biroli G. \and Bouchaud J.~P.} \REVIEW{EPL
  (Europhysics Letters)}{83}{2008}{46003}.

\bibitem{lechenault_epl2}
\Name{Lechenault F., Dauchot O., Biroli G. \and Bouchaud J.~P.} \REVIEW{EPL
  (Europhysics Letters)}{83}{2008}{46002}.

\bibitem{Majmudar_nature_2005}
\Name{Majmudar T.~S. \and Behringer R.~P.} \REVIEW{Nature}{435}{2005}{1079}.

\bibitem{ohernPRL01}
\Name{O'Hern C.~S., Langer S.~A., Liu A.~J. \and Nagel S.~R.} \REVIEW{Phys.
  Rev. Lett.}{86}{2001}{111}.

\bibitem{DH-book}
\Name{Berthier L., Biroli G., Bouchaud J.-P., Cipelletti L. \and Saarloos
  W.~V.} (Editors) \Book{Dynamical Heterogeneities in Glasses, Colloids, and
  Granular Media} (Oxford University Press) 2011.

\bibitem{PhysRevLett.104.165701}
\Name{Chaudhuri P., Berthier L. \and Sastry S.} \REVIEW{Phys. Rev.
  Lett.}{104}{2010}{165701}.

\bibitem{Ikeda:2012wp}
\Name{Ikeda A., Berthier L. \and Biroli G.} \REVIEW{arXiv eprint:
  1209.2814}{}{2012}{}.

\end{thebibliography}
\bibliographystyle{eplbib}
\end{document}